\documentclass[final,5p,times,twocolumn]{elsarticle}
\usepackage{xcolor}
\usepackage{amsmath,amssymb}
\usepackage{graphicx}
\usepackage{hyperref}
\usepackage{booktabs}

\journal{Physics Letters B}

\begin{document}

\begin{frontmatter}

\title{Nuclear matter and proton parton distributions in a light-front Hamiltonian framework }

\author[inst1,inst2]{Xiaoyi~Wu}
\ead{wuxiaoyi@impcas.ac.cn}

\author[inst1,inst2]{Sreeraj~Nair}
\ead{sreeraj@impcas.ac.cn}

\author[inst1,inst2]{Satvir~Kaur}
\ead{satvir@impcas.ac.cn}

\author[inst1,inst2]{Chandan~Mondal}
\ead{mondal@impcas.ac.cn}

\author[inst1,inst2,inst3]{Jiangshan~Lan}
\ead{jiangshanlan@impcas.ac.cn}

\author[inst1,inst2,inst3]{Xingbo~Zhao}
\ead{xbzhao@impcas.ac.cn}

\author[inst5]{J.~P.~B.~C.~de~Melo}
\ead{joao.mello@cruzeirodosul.edu.br}

\author[inst6]{Tobias~Frederico}
\ead{tobias@ita.br}

\address[inst1]{State Key Laboratory of Heavy Ion Science and Technology, Institute of Modern Physics, Chinese Academy of Sciences, Lanzhou 730000, China}
\address[inst2]{School of Nuclear Science and Technology, University of Chinese Academy of Sciences, Beijing 100049, China}
\address[inst3]{Advanced Energy Science and Technology Guangdong Laboratory, Huizhou, Guangdong 516000, China}
\address[inst5]{Laborat\'orio de F\'isica Te\'orica e Computacional--LFTC, Universidade Cruzeiro do Sul / Universidade Cidade de S\~ao Paulo, 015060-000, S\~ao Paulo, SP, Brazil}
\address[inst6]{Instituto Tecnol\'ogico de Aeron\'autica, 12.228-900, S\~ao Jos\'e dos Campos, SP, Brazil}

\begin{abstract}

We develop a light-front Hamiltonian formulation of symmetric nuclear matter within the quark–meson coupling model, using Basis Light-Front Quantization to solve the in-medium nucleon eigenvalue problem. The Hamiltonian incorporates confinement in the valence sector and is truncated to include up to one dynamical gluon. Medium effects are introduced via scalar and vector mean fields, yielding a self-consistent, density-dependent effective quark mass and modified nucleon structure. The resulting energy per nucleon, pressure, and incompressibility are consistent with empirical constraints at the saturation point. At nuclear saturation density, the gluon probability in the nucleon wave function increases slightly, while the valence probability and quark momentum fraction decrease. The unpolarized quark and gluon distributions show a noticeable enhancement at large momentum fraction ($x \gtrsim 0.4$), illustrated at an evolved scale of $Q^2 = 10 ~\mathrm{GeV}^2$.

\end{abstract}


\end{frontmatter}


\section{Introduction}

Understanding strongly interacting matter and the nuclear force directly from the quark and gluon degrees of freedom of Quantum Chromodynamics (QCD) remains a central challenge in nuclear theory~\cite{Aoki:2023qih}. Semi-phenomenological approaches play an important role in connecting QCD dynamics to nuclear observables. The quark--meson coupling (QMC) model~\cite{Guichon1988,Frederico1989,SaitoThomas1994} provides such a framework, where nucleons are described as composite quark systems interacting self-consistently with scalar and vector mean fields in the nuclear medium. This approach naturally incorporates medium modifications of nucleon structure and reproduces the saturation properties of symmetric nuclear matter (SNM).

Conventional implementations of the QMC model are formulated in equal-time quantization. In contrast, light-front (LF) Hamiltonian dynamics~\cite{BrodskyPauliPinsky1998} offers a framework in which hadronic structure is described in terms of light-front wave functions (LFWFs), with a direct connection to parton distribution functions (PDFs)~\cite{Bakker:2013cea}. Basis Light-Front Quantization (BLFQ)~\cite{Vary:2009gt,Zhao2014,Vary2025BLFQ} provides a nonperturbative realization of this approach, enabling the computation of hadronic bound states from a LF Hamiltonian.

In this work, we formulate the equation of state (EOS) of symmetric nuclear matter within a realization of the QMC model in which the nucleon is described as an eigenstate of an in-medium LF Hamiltonian truncated to include the valence three-quark sector $|qqq\rangle$ and the sector with one dynamical gluon $|qqqg\rangle$~\cite{Vary2025BLFQ}. The nucleon is embedded in the nuclear medium and reacts at the quark--gluon level through the coupling of quarks to scalar and vector mean fields, which are the adopted nuclear interactions in the framework of Quantum Hadrodynamics (QHD) for the in-medium interaction between nucleons in order to explore the EOS for SNM (see the earlier works~\cite{Guichon1988,Frederico1989}). This leads to a self-consistent equation for the scalar mean field and to a density-dependent nucleon mass obtained from the LF Hamiltonian eigenvalue problem. We compute the energy per nucleon, pressure, and incompressibility of SNM, and compare the results with empirical constraints and conventional equal-time QMC approaches.

The LF formulation further allows direct access to parton distributions from the nucleon wave function defined in the LF Fock space up to one dynamical gluon. We therefore compute the in-medium quark and gluon PDFs and analyze their density dependence in SNM at the nucleon scale. As an illustrative example, results are also presented after scale evolution to $Q^2 = 10\,\mathrm{GeV}^2$, in order to identify the characteristic mean-field modifications before and after scale evolution.

The paper is organized as follows. In Sec.~\ref{secLFH}, we review the LF Hamiltonian formulation for the nucleon with confinement in the valence sector and truncation up to one dynamical gluon~\cite{Vary2025BLFQ}, and outline its implementation within the BLFQ framework~\cite{Vary:2009gt}. In Sec.~\ref{sec:LFHSNM}, we present the in-medium LF Hamiltonian, where quarks interact with scalar and vector mean fields, leading to a density-dependent nucleon mass and wave function. In Sec.~\ref{sec:QHD}, we introduce the QHD mean-field equations for symmetric nuclear matter and determine the model parameters by fitting the energy per nucleon at saturation. In Sec.~\ref{sec:SNM_EOS}, we present the resulting EOS for SNM at zero temperature, including the energy per nucleon and pressure, and compare with experimental constraints and other models. In Sect.~\ref{sec:PDF_SNM}, we analyze the density dependence of the leading-twist quark and gluon momentum distributions at the nucleon scale and after evolution to $10\,\mathrm{GeV}^2$. In Sec.~\ref{sec:summary}, we summarize our results and discuss prospects for future applications of the model.

\section{Light-front Hamiltonian formulation}\label{secLFH}
In the LF framework~\cite{Dirac:1949cp, BrodskyPauliPinsky1998}, dynamics are formulated in terms of LF variables: $v^\pm = v^0 \pm v^3$ and $\mathbf{v}_\perp=(v_1,\,v_2)$.
One of the key advantages of this formalism is that it enables a Hamiltonian description of relativistic systems with a well-defined Fock space expansion of states. In this framework, at
fixed LF time, the proton state with momentum $P$ and LF helicity $\Lambda$ can be expanded in terms of quark and gluon degrees of freedom as
\begin{equation}
|P,\Lambda\rangle = \psi_{\lbrace qqq \rbrace} |qqq\rangle + \psi_{\lbrace qqqg \rbrace} |qqqg\rangle + \psi_{\lbrace qqqq\bar{q}\rbrace} |qqqq\bar{q}\rangle + \cdots,
\end{equation}
where $|qqq\rangle$ denotes the valence three-quark component, $|qqqg\rangle$ represents the three-quark plus one gluon state, and $|qqqq\bar{q}\rangle$ corresponds to higher Fock sectors involving sea quark-antiquark pairs. The coefficients $\psi_n$ encode the corresponding LFWFs. In practice, the infinite Fock space expansion must be truncated; in this work, we retain only the lowest two Fock sectors, namely the valence $|qqq\rangle$ and the $|qqqg\rangle$ components.

The generators of spacetime translations are given by the LF momenta $P^\pm = P^0 \pm P^3$ and $\mathbf{P}_\perp$, with the LF Hamiltonian $P^-$. The eigenvalue problem for a bound state with invariant mass $M$ can be expressed as~\cite{BrodskyPauliPinsky1998}
\begin{equation}
P^- |P,\Lambda \rangle = \frac{M^2 + P_\perp^2}{P^+} |P,\Lambda\rangle,
\end{equation}
which provides the starting point for nonperturbative approaches to nucleon structure.

To render the LF eigenvalue problem tractable within a truncated Fock space, we employ an effective LF Hamiltonian that captures the dominant quark-gluon dynamics relevant for the proton. Working in the LF gauge $A^+ = 0$, the effective Hamiltonian, expressed in terms of the dynamical quark and gluon fields, is separated into kinetic, interaction, and confining components~\cite{Xu:2022yxb, Lin:2023ezw, Yu:2024mxo, Zhu:2024awq, Lin:2024ijo, Zhang:2025nll, Nair:2025sfr, Zhang:2026dzi, Zhang:2026cig},
\begin{equation}\label{eq:HLFeff}
P^-_{\mathrm{eff}} = P^-_{\mathrm{kin}} + P^-_{\mathrm{int}} + P^-_{\mathrm{conf}}.
\end{equation}

The kinetic terms for quarks and gluons are given by~\cite{Xu:2022yxb}
\begin{equation}
\hspace{-.2cm}
P^-_{\mathrm{kin}} = \hspace{-.1cm}\int dx^- d^2x_\perp \left[
\bar{\varphi} \gamma^+ \frac{m_q^2 + (i\partial_\perp)^2}{2 i \partial^+} \varphi
+  A_a^\mu \frac{m_g^2 + (i\partial_\perp)^2}{2} A^a_\mu
\right],
\end{equation}
where $\varphi$ and $A^\mu_a$ denote the quark and gluon fields, respectively, $m_q$ is the effective quark mass, and $m_g$ is a phenomenological gluon mass introduced to model nonperturbative effects. 

The interaction part includes the quark-gluon coupling as well as instantaneous interactions characteristic of LF dynamics~\cite{BrodskyPauliPinsky1998, Xu:2022yxb},
\begin{equation}
\begin{aligned}
P^-_{\mathrm{int}} &= g_s \int dx^- d^2x_\perp \, \bar{\varphi} \gamma_\mu T^a A_a^\mu \varphi
\\
&+ \frac{g_s^2}{2} \int dx^- d^2x_\perp \,
\bar{\varphi} \gamma^+ T^a \varphi \, \frac{1}{(i\partial^+)^2} \, \bar{\varphi} \gamma^+ T^a \varphi.
\end{aligned}
\end{equation}
Here, $g_s$ is the strong coupling constant and $T^a = \lambda^a/2$ are the generators of SU(3) color.

Since higher Fock sector contributions are not explicitly included in the present truncation, their effects are incorporated through effective parameters and mass renormalization. In particular, we introduce a quark mass counterterm defined as
$\delta m_q = m_0 - m_q,$
where $m_0$ denotes the bare quark mass and $m_q$ is the effective (renormalized) quark mass used in the Hamiltonian. This counterterm is included in the leading Fock sector to account for self-energy corrections arising from quantum fluctuations~\cite{Xu:2022yxb}. In addition, we allow for a separate quark mass parameter in the interaction vertex, denoted by $m_f$, which is treated independently from the kinetic mass~\cite{Glazek:1992aq, Xu:2022yxb}. This provides additional flexibility in modeling the effective quark-gluon coupling within the truncated framework. This parameter also influences helicity-changing contributions in the LFWF, thereby affecting the spin structure of the proton.

In addition to the kinetic and interaction terms, we introduce an effective confinement contribution, denoted by $P^-_{\mathrm{conf}}$, which acts only within the valence Fock sector. This term is implemented through a two-body operator that depends on the relative transverse coordinates of the constituents and is motivated by LF holographic considerations~\cite{Brodsky:2014yha}. Its explicit form can be written as~\cite{Li:2015zda}
\begin{equation}
P^+P^-_{\mathrm{conf}} = \frac{\kappa^4}{2} \sum_{i \neq j} \left [ {\bf r}^2_{ij\perp}-\frac{\partial_{x_i} \left( x_i x_j \, \partial_{x_j} \right)}{(m_i + m_j)^2} \right ],
\end{equation}
where the transverse separation between partons is defined as $\mathbf{r}_{ij\perp} = \sqrt{x_i x_j}\, (\mathbf{r}_{i\perp} - \mathbf{r}_{j\perp})$, which is closely related to the holographic variable. Here, $\kappa$ sets the overall confinement scale, and $\partial_{\mathbf{r}_{ij\perp}}$ denotes differentiation with respect to the relative transverse coordinate.

We emphasize that this confining interaction is restricted to the valence sector. For higher Fock components such as $|qqqg\rangle$, no explicit confinement term is included. Instead, confinement effects in these sectors are effectively accounted for through the restricted basis in the transverse direction and the introduction of an effective gluon mass.

Within the BLFQ framework~\cite{Vary:2009gt}, the basis states in each Fock sector are constructed as direct products of single-particle states, $|\alpha\rangle = \otimes_i |\alpha_i\rangle$ (for the recent review, see~\cite{Vary2025BLFQ}). In the longitudinal direction, each constituent is confined within a one-dimensional box of length $2L$. Antiperiodic boundary conditions are imposed for fermions, while periodic boundary conditions are used for bosons. As a result, the longitudinal momentum is discretized according to $p_i^+ = \frac{2\pi}{L} \, k_i,$
where $k_i$ takes half-integer values for fermions and integer values for bosons. The zero mode for bosonic degrees of freedom is excluded in this work.

In the transverse direction, we employ a two-dimensional harmonic oscillator (HO) basis characterized by quantum numbers $(n_i, m_i)$ and a scale parameter $b$, i.e., $\phi_{n_i,m_i} ({\bf p}_{\perp i}; b)$~\cite{Zhao:2014xaa}. Each single-particle state is specified by the set $|\alpha_i\rangle = |k_i, n_i, m_i, \lambda_i\rangle$, where $\lambda_i$ denotes the LF helicity. The total angular momentum projection is conserved and satisfies $\sum_i (m_i + \lambda_i) = \Lambda$.

To render the basis finite, we introduce truncations in both longitudinal and transverse directions~\cite{Zhao:2014xaa}. The total longitudinal momentum is fixed by $K = \sum_i k_i$, such that the momentum fraction is given by $x_i = k_i / K$. In the transverse plane, the basis is limited by the condition $\sum_i (2n_i + |m_i| + 1) \leq N_{\max}$. These truncations effectively regulate both infrared and ultraviolet scales, $\Lambda_{\rm IR} \approx b/\sqrt{N_{\rm max}}$ and $\Lambda_{\rm UV} \approx b\sqrt{N_{\rm max}}$, of the system.

Solving the resulting LF Hamiltonian eigenvalue problem yields the proton LFWFs, which can be expressed as an expansion in the chosen basis,
\begin{equation}
\Psi^{\Lambda}_{\mathcal{N}, \lbrace \lambda_{i} \rbrace}(\{x_i, \mathbf{p}_{i\perp}\}) = \sum_{\{n_i, m_i\}} \psi^{\Lambda}_\mathcal{N}(\{\alpha_i\}) \prod_i^\mathcal{N} \phi_{n_i, m_i}(\mathbf{p}_{i\perp}; b),
\end{equation}
for the $\mathcal{N}-$th particle Fock-component, and the  $\phi_{n_i, m_i}$ denote the transverse HO basis functions and $\psi^{\Lambda}_\mathcal{N}(\{\alpha_i\})$ represent the longitudinal and spin-dependent components of the eigenvectors. The solutions receive contributions from both the valence sector $|qqq\rangle$ and the next Fock sector $|qqqg\rangle$, corresponding to $\mathcal{N}=3$ and $\mathcal{N}=4$ constituents, respectively. We note that the $|qqqg\rangle$ sector admits two color-singlet configurations, requiring an additional label to fully specify the states.

In our numerical implementation, we adopt a finite basis characterized by the truncation parameters $\{N_{\max}, K\} = \{9, 16.5\}$, following the strategy outlined in Ref.~\cite{Xu:2022yxb}. The transverse dynamics are described using a HO basis with the scale parameter $b = 0.7~\mathrm{GeV}$, while the instantaneous interaction is regulated by an ultraviolet cutoff $b_{\mathrm{inst}} = 3~\mathrm{GeV}$. The model parameters, including the constituent quark masses, effective gluon mass, confinement strength, and couplings, are chosen as $\{m_u, m_d, m_g, \kappa, m_f, g_s\} = \{0.31, 0.25, 0.50, 0.54, 1.80, 2.40\}$ in units of GeV (except for $g_s$), and are fixed by fitting to the proton mass and its electromagnetic properties in the vacuum~\cite{Xu:2022yxb}.  Noteworthy to mention the special dynamical role played by the quark mass in the vertex $m_f$, which normalizes the spin-flip matrix element of the quark-gluon interaction, necessary to split the $\pi$ and the $\rho-$meson masses in the BLFQ framework~\cite{Lan:2024ais,Lan:2021wok,Xu:2022yxb}.

At the chosen model scale, the normalization of the wavefunction indicates that the proton is composed of approximately $42\%$ valence ($|qqq\rangle$) and $58\%$ quark-gluon ($|qqqg\rangle$) contributions, highlighting the significant role of gluonic degrees of freedom in the proton structure.

\section{In-medium effective light-front Hamiltonian}\label{sec:LFHSNM}

In a QMC-like model, the nuclear medium is described at the mean-field level by classical scalar and vector fields, $\sigma$ and $\omega^\mu$, generated self-consistently by the surrounding nuclear matter~\cite{Guichon1988,Frederico1989,SaitoThomas1994}. In uniform and isospin-SNM, only the time component $\omega^0$ of the vector field survives, while the scalar field modifies the effective (renormalized) quark mass. Our goal is to embed the intrinsic effective LF Hamiltonian, Eq.~\eqref{eq:HLFeff}, in the nuclear medium, where the constituent quarks couple to the mean fields $\sigma$ and $\omega_0$, thereby describing the response of the nucleon to the nuclear environment.

The in-medium nucleon is an eigenstate of the effective LF Hamiltonian coupled to the mean fields,
\begin{equation}\label{eq:P-*}
P_{\rm eff}^{-*}
=
P_{\rm eff}^-
+
\int \hspace{-.1cm}dx^- d^2x_\perp \,
\bar{\varphi}\,\gamma^+\,
\frac{(m_q-g_\sigma^q \sigma)^2-m_q^2}{2i\partial^+}\,
\varphi
+3g_\omega^q\omega_0,
\end{equation}
where $P_{\rm eff}^-$ is the vacuum LF Hamiltonian used to construct the nucleon in the truncated Fock space $|qqq\rangle + |qqqg\rangle$. The quark coupling constants to the scalar and vector isoscalar fields are denoted by $g_\sigma^q$ and $g_\omega^q$, respectively. In the rest frame of isotropic SNM, only the time component of the vector field is nonzero ($\omega^-=\omega^0$). In addition, as is standard in QMC models, we assume that the nucleon moves sufficiently slowly with respect to the nuclear matter so that only the time component of the vector mean field contributes in Eq.~\eqref{eq:P-*}. In fact, the mean-field equation for SNM depends only on the effective nucleon mass, while the contribution of the vector mean field enters through the SNM equation of state. 

The scalar mean field shifts the quark mass in the nuclear medium according to
\begin{equation}
m_q^* = m_q - g_\sigma^q \sigma \,,
\end{equation}
while the vector field produces an overall shift in the nucleon LF energy, as given by the last term in Eq.~\eqref{eq:P-*}. The corresponding nucleon coupling to the vector field is $g_\omega = 3g_\omega^q$. Furthermore, the vertex mass parameter $m_f$ in $P^-_\text{eff}$ is assumed to be shifted as
\begin{equation}
m_f^* = m_f - g_\sigma^q \sigma \,,
\end{equation}
which gives only a marginal modification because of its large vacuum value.

The coupling to the scalar field modifies the internal dynamics of the nucleon, since the LF kinetic-energy operator depends explicitly on $m_q^*$, as seen in Eq.~\eqref{eq:P-*}. Consequently, the BLFQ eigenvalue problem in nuclear matter becomes
\begin{equation}\label{eq:p*omega}
\left(P_{\rm eff}^{-*}-g_\omega\omega_0\right)
|\Psi_{\mathcal{N},\{\lambda_i\}}^*\rangle
=
\frac{\left[M_N^*(\sigma)\right]^2+P_\perp^2}{P^+}
|\Psi_{\mathcal{N},\{\lambda_i\}}^*\rangle ,
\end{equation}
which yields a density-dependent nucleon mass $M_N^*(\sigma)$ through the coupling of the quarks to the scalar mean field, as well as medium-modified LFWFs and, consequently, modified parton distributions.

The nucleon--$\sigma$ coupling constant is obtained from the Feynman--Hellmann theorem and is given by~\cite{Batista2002}
{
\begin{equation}\label{eq:gsN}
g^N_\sigma = -\frac{\partial M_N^*}{\partial \sigma} \,.
\end{equation} }
This coupling is density dependent and therefore affects the EOS, as expected in QMC models. One consequence is the softening of the EOS near the saturation point.

\section{QHD mean-field equations for symmetric nuclear matter}\label{sec:QHD}

{ Within QHD~\cite{Serot:1984ey}, the energy density and pressure of symmetric nuclear matter are obtained from the diagonal components of the energy--momentum tensor evaluated at the mean-field level. The energy density is given by}
\begin{equation}\label{eq:Edens}
	\mathcal{E} =
	\frac{\gamma}{(2\pi)^3}
	\int_{|\mathbf{k}|<k_F} d^3k \,
	\sqrt{\mathbf{k}^2 + M_N^{*2}}
	+ \frac{1}{2} m_\sigma^2 \sigma^2
	+ \frac{1}{2} m_\omega^2 \omega_0^2 \,,
\end{equation}
while the pressure is
\begin{equation}
	\mathcal{P} =
	\frac{\gamma}{3(2\pi)^3}
	\int_{|\mathbf{k}|<k_F} d^3k \,
	\frac{\mathbf{k}^2}{\sqrt{\mathbf{k}^2 + M_N^{*2}}}
	- \frac{1}{2} m_\sigma^2 \sigma^2
	+ \frac{1}{2} m_\omega^2 \omega_0^2 \,,
\end{equation}
where $k_F$ is the Fermi momentum and $\gamma=4$ is the spin-isospin degeneracy factor for SNM.

\begin{figure}[t]
	\centering
	\includegraphics[width=8cm]{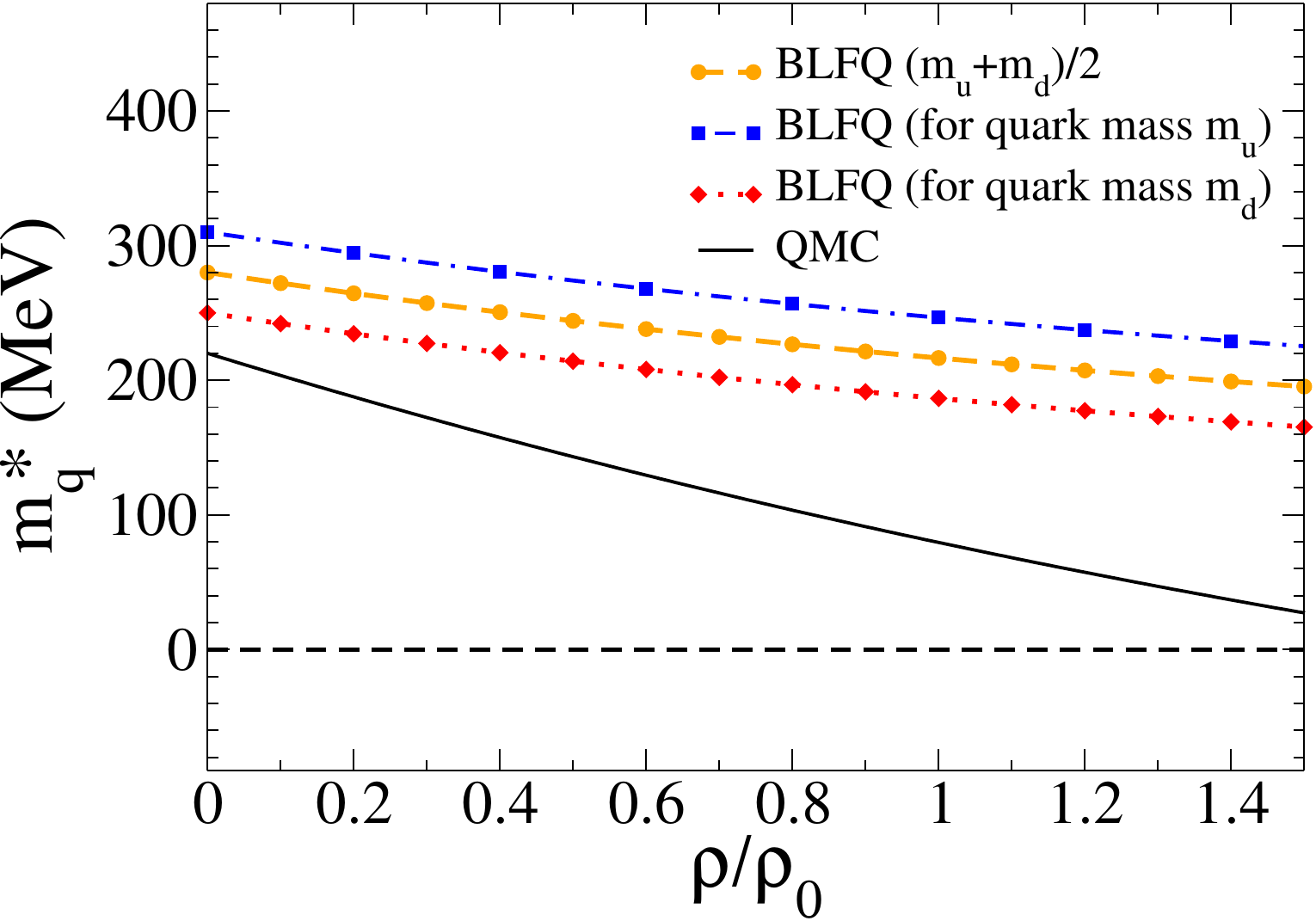}
    \hspace{1cm}
    \includegraphics[width=8cm]
    {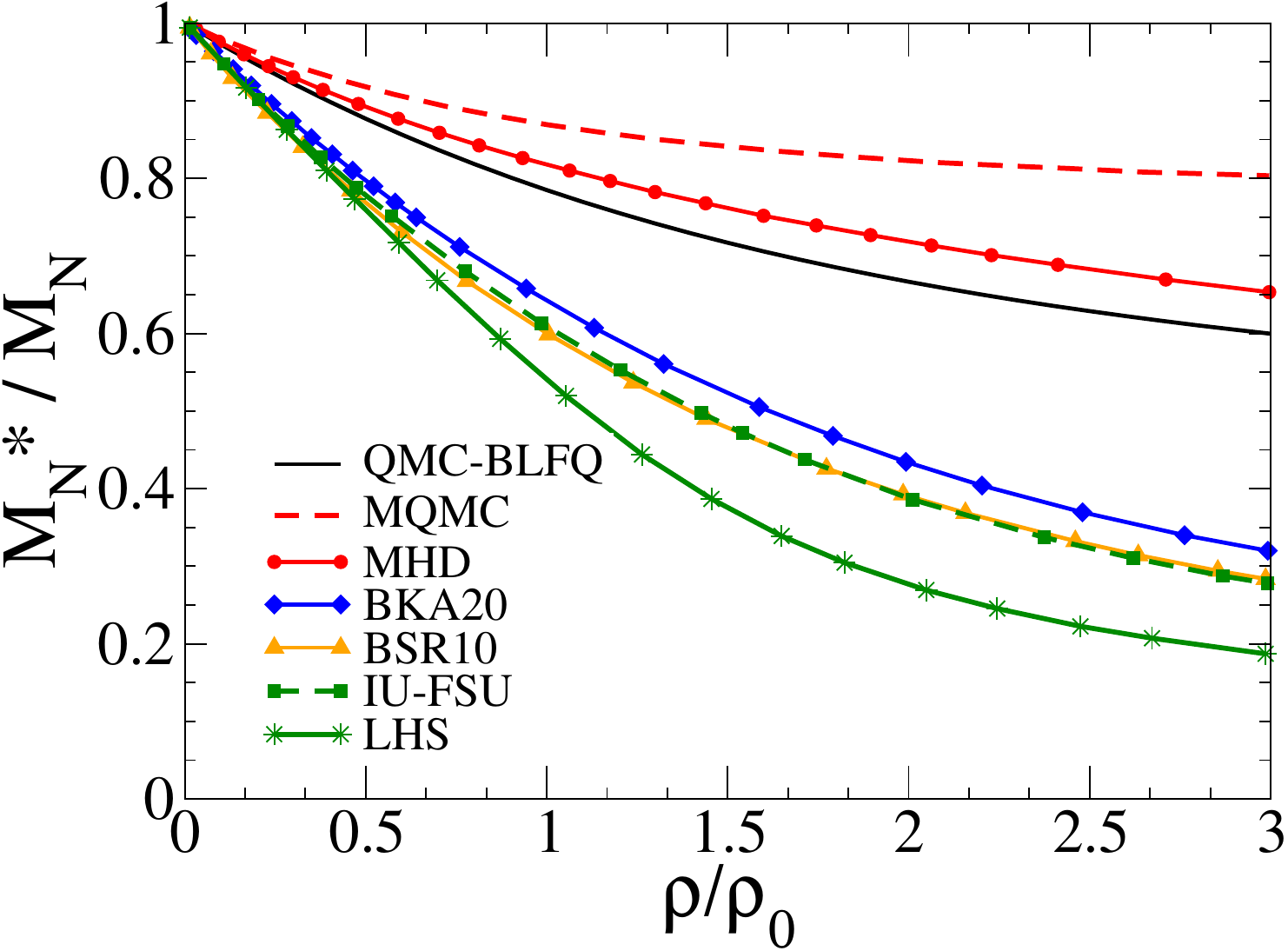}
	\caption{{\it Top.} Quark masses as a function of $\rho/\rho_0$ for the QMC-BLFQ model compared with the QMC model~\cite{Yabusaki:2023zin}, where the $u$ and $d$ quarks are degenerate. {\it Bottom.} Ratio $M_N^*/M_N$ as a function of $\rho/\rho_0$ for the QMC-BLFQ model, compared with the Walecka model (LHS)~\cite{Reinhard1989}, medium-modified holographic hadron dynamics (MHD)~\cite{dePaula:2020bte}, the modified quark--meson coupling model (MQMC)~\cite{Batista2002}, and relativistic mean-field parametrizations BKA20~\cite{Agrawal:2010wg}, BSR10~\cite{Dhiman2007}, and IU-FSU~\cite{Fattoyev2010}.}
	\label{fig:mqstarv1}
\end{figure}


\begin{figure*}[t]
	\centering
	\includegraphics[width=8cm]
    {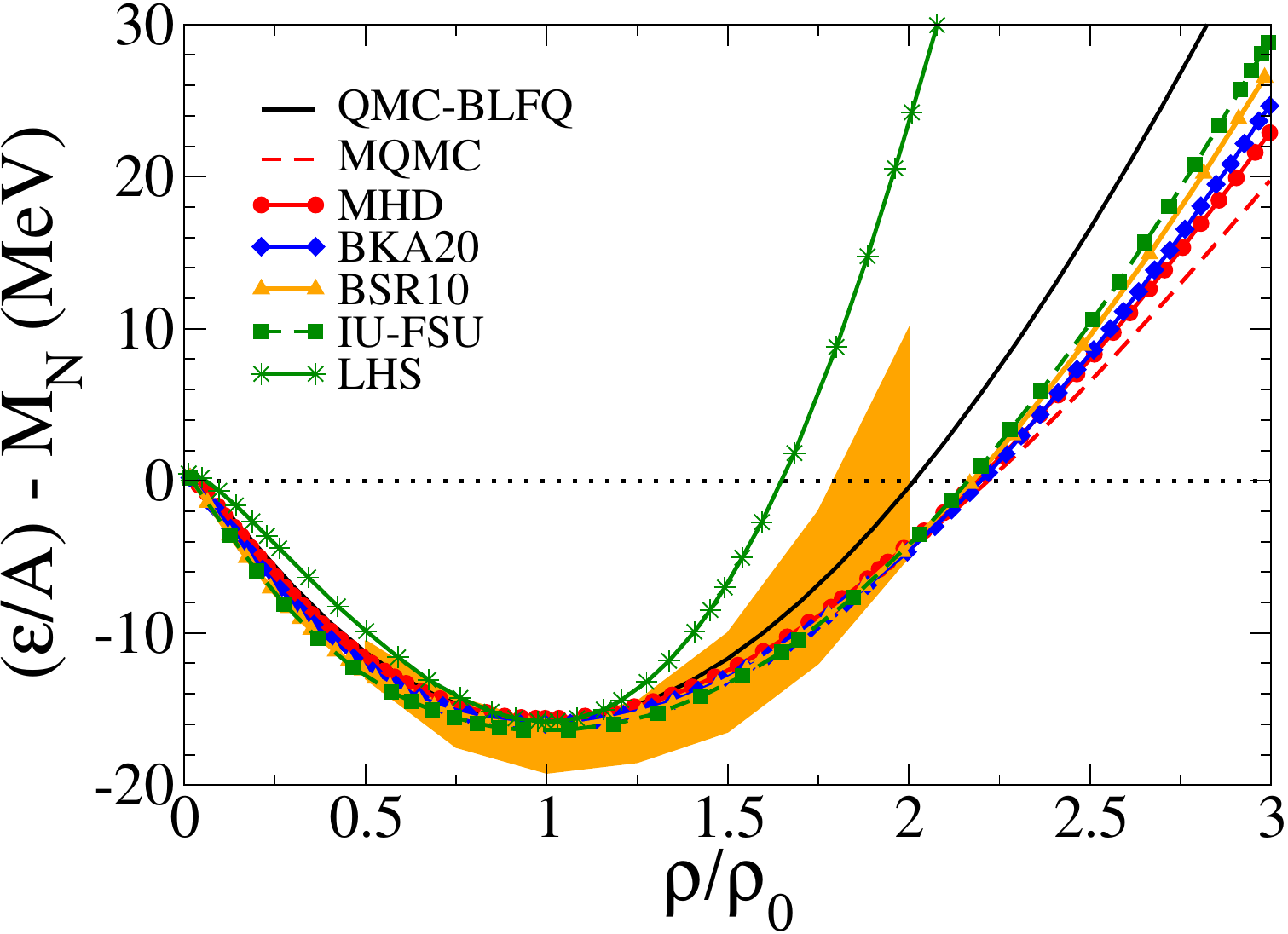}
    \hspace{1cm}
\includegraphics[width=8cm]
    {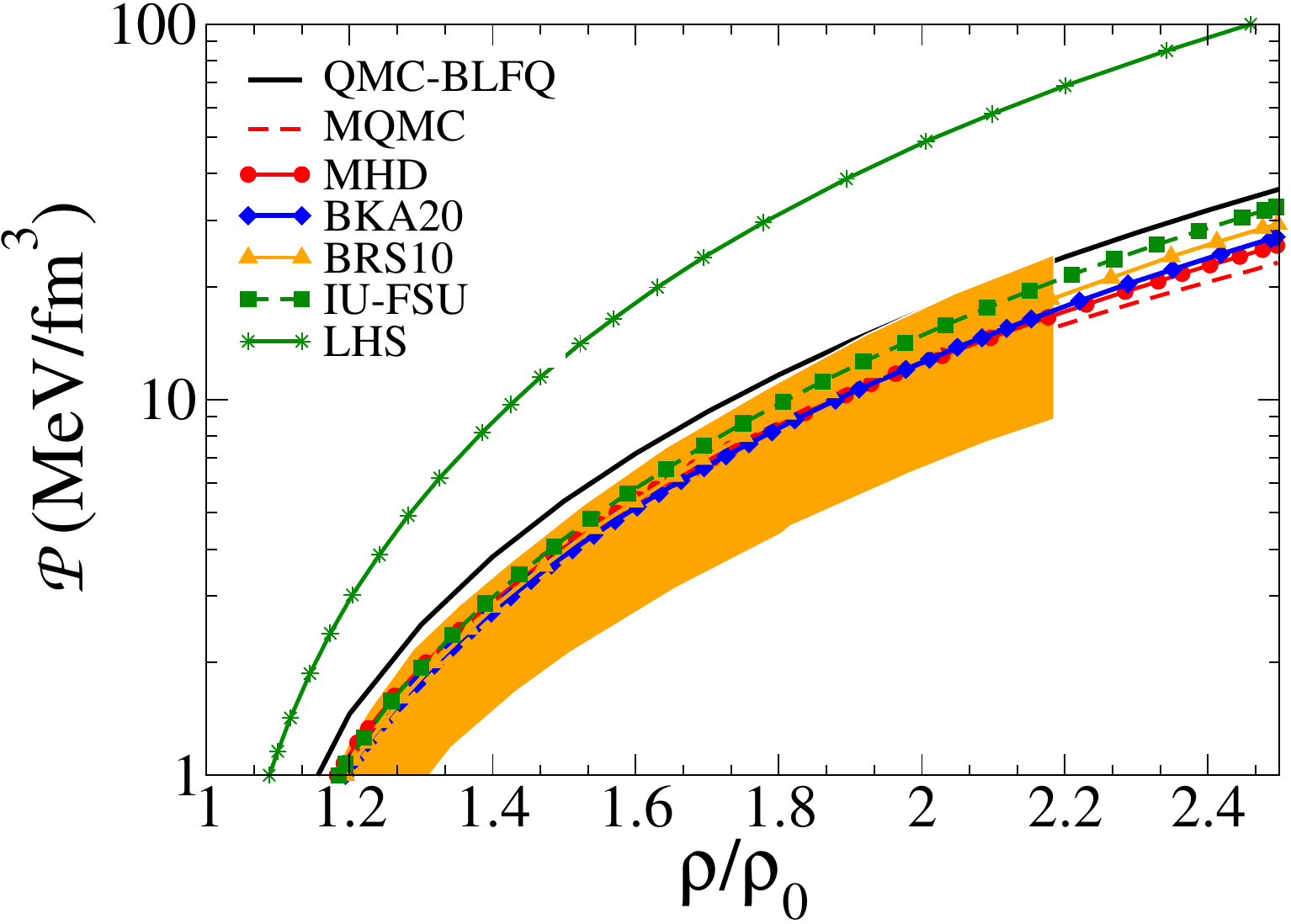}

\includegraphics[width=8cm]
    {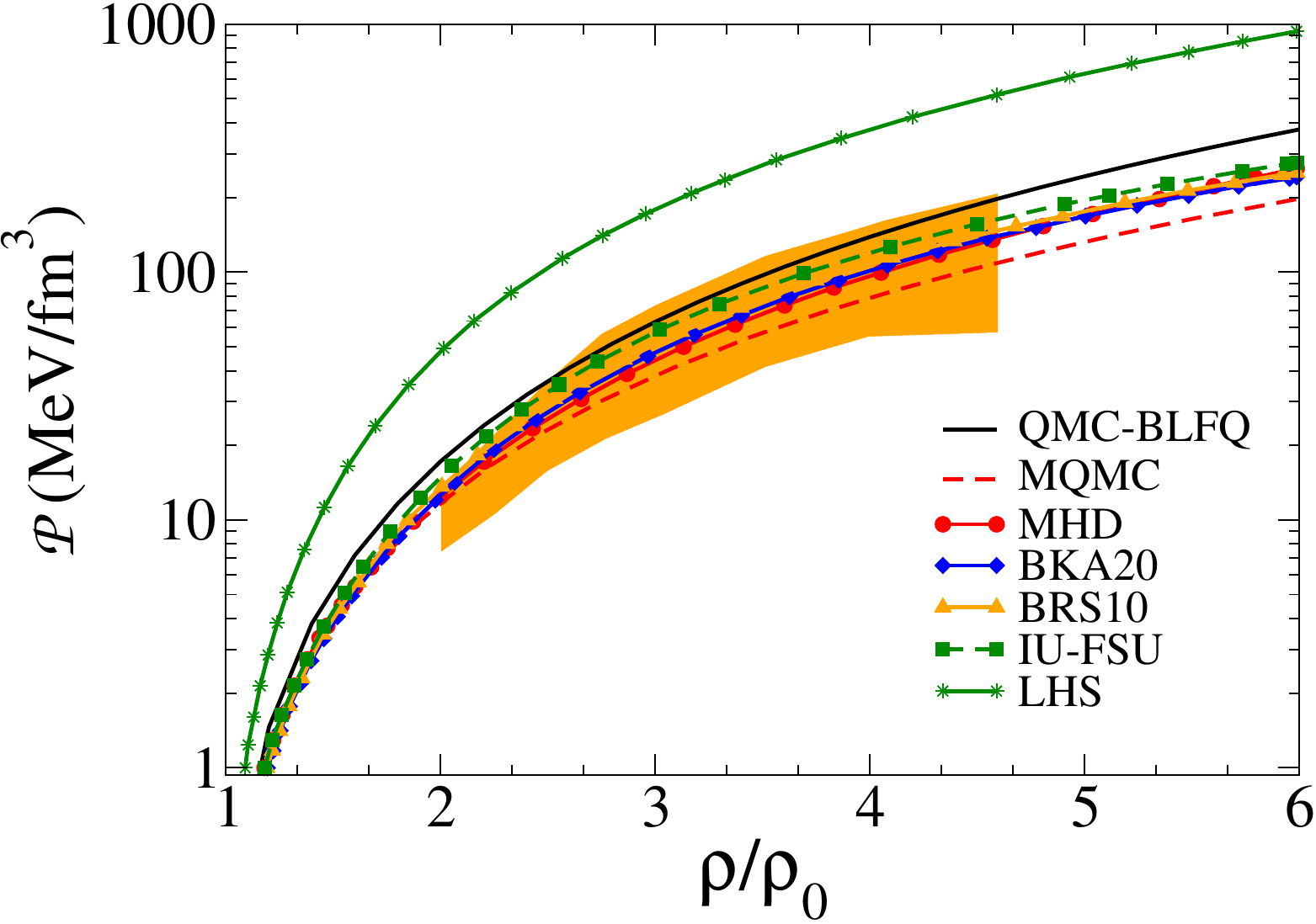}
    \hspace{1cm}
\includegraphics[width=8cm]
    {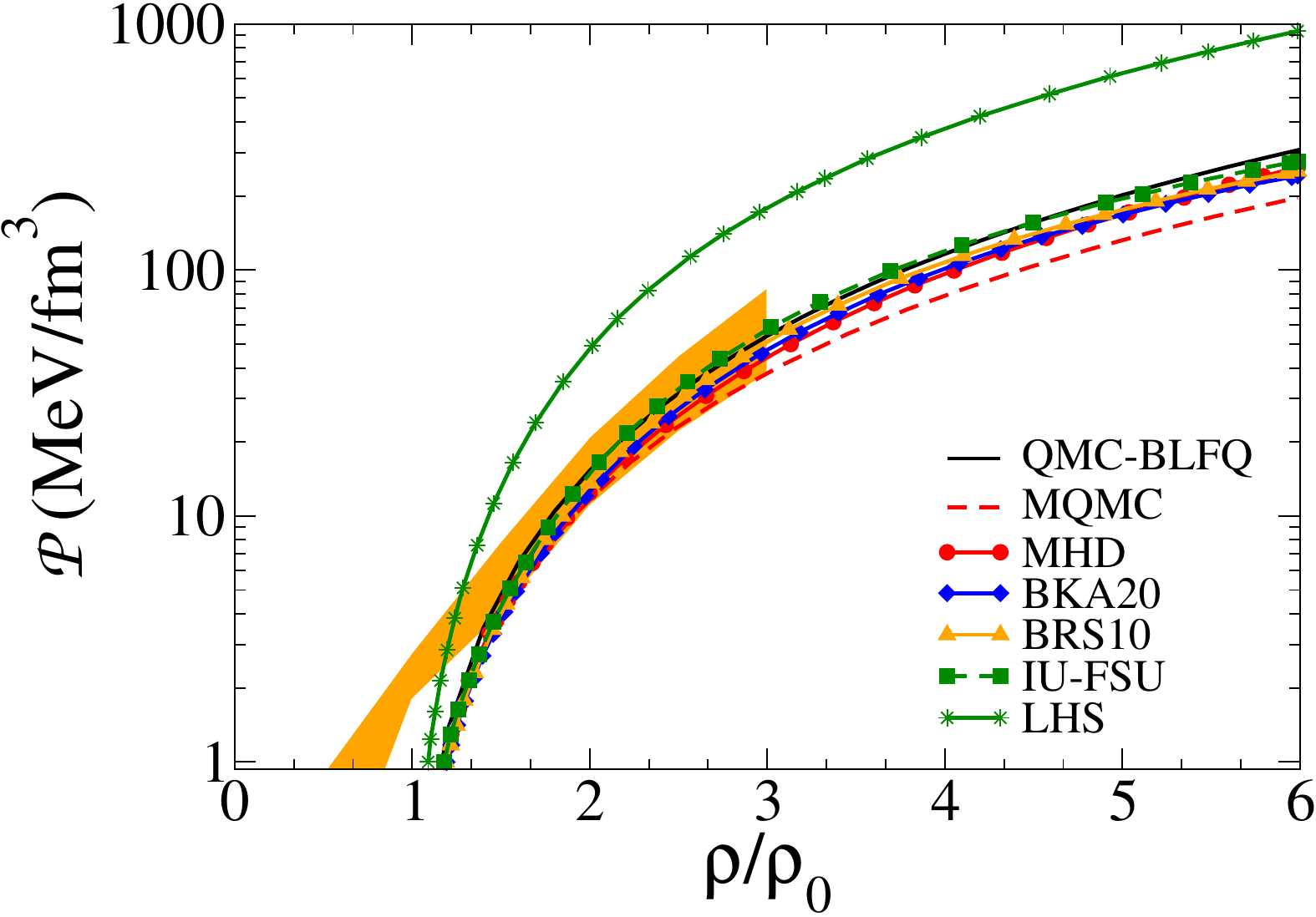}
    
	\caption{ Energy per nucleon and pressure vs. $\rho/\rho_0$ for the present  QMC-BLFQ model and other models/parametrizations:  LHS~\cite{Reinhard1989}, MHD~\cite{dePaula:2020bte}, BKA20~\cite{Agrawal:2010wg},  BSR10~\cite{Dhiman2007} and IU-FSU~\cite{Fattoyev2010}.  Bands: data extracted from  ~\cite{Huth2021,LeFevre2015,Russotto2016} (top left), \cite{Lynch2009} (top right),  \cite{Danielewicz:2002pu} (bottom left), \cite{Huth2021} and references therein (bottom right).}
	\label{densi}
\end{figure*}

The self-consistent equation for the scalar field follows from minimizing the energy density in Eq.~\eqref{eq:Edens} with respect to $\sigma$, which gives~\cite{Batista2002}:
\begin{equation}\label{eq:sigmaself}
	\sigma =
	-\frac{\partial M_N^*}{\partial \sigma}
	\frac{1}{m_\sigma^2}
	\frac{\gamma}{(2\pi)^3}
	\int_{|\mathbf{k}|<k_F} d^3k \,
	\frac{M_N^*}{\sqrt{\mathbf{k}^2 + M_N^{*2}}} \,.
\end{equation}
Compared with the linear Walecka $\sigma$--$\omega$ model, in which $M_N^*=M_N-g^N_\sigma \sigma$ with constant $g^N_\sigma$ (see, e.g.,~\cite{Serot:1984ey}), the present approach contains a nontrivial dependence of the nucleon effective mass on the scalar mean field, together with a density-dependent nucleon--$\sigma$ coupling. The coupling $g^N_\sigma(\sigma)$ in Eq.~\eqref{eq:gsN} depends on the scalar mean field and, consequently, acquires a density dependence, which is determined self-consistently through the solution of Eq.~\eqref{eq:sigmaself}. This dependence of the nucleon properties on $\sigma$ is determined by the underlying quark--gluon dynamics in nuclear matter through the in-medium effective LF Hamiltonian, Eq.~\eqref{eq:P-*}.

The above self-consistent equation for the scalar field and the one for the vector field can be written compactly as: 
\begin{equation}
\sigma = \frac{g^N_\sigma}{m_\sigma^2} \rho_s
\quad{\rm and}\quad
\omega_0 = \frac{g^N_\omega}{m_\omega^2} \rho ,
\end{equation}
where 
\begin{equation}
\rho_s=\frac{\gamma}{(2\pi)^3}\int_{|\mathbf{k}|<k_F} d^3k \,\frac{M^*_N}{\sqrt{\mathbf{k}^2+M^{*2}_N}} \quad{\rm and}\quad \rho=\frac{\gamma}{6\pi^2}k^3_{F}\,,
\end{equation}
denote the scalar and baryon densities, respectively. The parameters $g^q_{\sigma}$ and $g^q_{\omega}/m_\omega$ are fitted to reproduce the properties of SNM at saturation density, with the scalar meson mass fixed at $m_\sigma = 550$ MeV, consistent with previous QMC studies~\cite{Batista2002,dePaula:2020bte}.
The adopted value of the binding energy per nucleon, defined as $(B=\mathcal{E}/A-M_N)$, at the saturation density $\rho=\rho_{0}=0.15$ fm$^{-3}$ is $B_{0}=-15.8$ MeV~\cite{Reinhard1989}. This fitting procedure yields the parameters $g^q_{\sigma}=2.6445$ and $g^q_{\omega}/m_\omega=0.0039$ MeV$^{-1}$.

The incompressibility is given by
\begin{equation}
K = 9\,\frac{\partial \mathcal{P}}{\partial \rho}\,.
\label{incom}
\end{equation}
For SNM a recent analysis provides  $K_0\equiv K (\rho_0 )=220\pm 40\,$MeV~\cite{Wang2018}.
For the linear Walecka model (LHS)~\cite{Reinhard1989} fitted for the saturation parameters $\rho_0$ and $B_0$ the result is $K_0 = 548\,$MeV,
well above the value extracted from the recent analysis of the rapidity-dependent elliptic flow in heavy-ion collisions~\cite{Wang2018}, while $K_0=313.7\,$MeV for the present parametrization of the QMC-BLFQ model.

\section{Symmetric nuclear matter at zero temperature}\label{sec:SNM_EOS}

This section is devoted to the study of the SNM EOS provided by the QMC model within the BLFQ framework (QMC-BLFQ) with the parameters obtained from fitting the saturation density and energy per nucleon. In addition,  we discuss the in-medium quark and nucleon masses as a function of the nuclear density and compare to other models.

We begin by analyzing the  QMC-BLFQ results for the in-medium quark mass in
Fig.~\ref{fig:mqstarv1} (top), which shows the effective quark masses obtained within two different frameworks: the QMC model~\cite{Yabusaki:2023zin}
and the present approach, QMC-BLFQ, that combines the quarks coupled to the scalar and vector fields with the nucleon internal dynamics. The nucleon is an eigenstate of the in-medium LF Hamiltonian given by Eq.~\eqref{eq:P-*}, which is obtained simultaneously with the self-consistent  solution of Eq.~\eqref{eq:sigmaself} for the scalar mean field. The advantage  of LF Hamiltonian approach is the microscopic dynamics that defines the nucleon state beyond the valence sector, where it takes into account Fock components including gluons, and potentially sea quarks. In contrast, the QMC model assumes isospin symmetry and deals only with the valence nucleon state in the bag model. Qualitatively, it is seen that in both models in-medium quark masses drop with the density, with the QMC-BLFQ more resilient than QMC to the density changes, but enough to provide a reasonable SNM incompressibility at the saturation point.

Figure~\ref{fig:mqstarv1} (bottom) shows the dependence of $M_N^*/M_N$ with the nuclear density, where the results for the QMC-BLFQ model  are compared to others: linear Walecka model (LHS)~\cite{Reinhard1989}; Quantum Hadrodynamics (QHD)  models from references BKA20~\cite{Agrawal:2010wg}, BSR10~\cite{Dhiman2007}, and IU-FSU~\cite{Fattoyev2010}; medium modified holographic hadron dynamics (MHD)~\cite{dePaula:2020bte}; and   modified QMC model (MQMC)~\cite{Batista2002} parametrized to reproduce the same values of $\rho_0$, $B_0$ and $K_0$ as in the MHD model. It is worth mentioning that in the dependence of  $M^*_N/M_N$ on $\rho/\rho_0$, the nucleon is considered to be clustered for the QMC-BLFQ, MHD and MQMC models, as well as for the LHS and QHD. The models where the nucleon has an internal structure exhibit similar behavior with a softer dependence of $M^*_N/M_N$ with the density compared to those where the nucleon is point-like.

The EOS of symmetric nuclear matter is illustrated in 
Fig.~\ref{densi}, where we present the energy per nucleon and pressure as a function of the nuclear 
density. In the top left panel, the energy per nucleon from QMC-BLFQ shows a good agreement with the band corresponding to the  data extracted from Refs.~\cite{Huth2021,LeFevre2015,Russotto2016}.  Comparing with QHD, MHD and MQMC the EOS of QMC-BLFQ is somewhat harder, but not as hard as that of LHS.

\begin{table*}[tbh]
\centering
\begin{minipage}{0.75\textwidth}
\caption{Moments $\langle x^n\rangle_{q,g}$ with $n=0,1$ in the nuclear medium as a function of $\rho/\rho_0$ for the proton at the nucleon scale $\mu_0^2=0.291~\mathrm{GeV}^2$. Here, ``tot'' denotes the sum of the contributions from the $|qqq\rangle\,(3q)$ and $|qqqg\rangle\,(3qg)$ sectors. For the neutron, the corresponding results are obtained by interchanging $u$ and $d$. 
}
\label{tab:pdf_moments}
\centering
\begin{tabular}{c ccc ccc c}
\toprule
& \multicolumn{3}{c}{$u$ quark} & \multicolumn{3}{c}{$d$ quark} & gluon \\
\cmidrule(lr){2-4} \cmidrule(lr){5-7}
$\rho/\rho_0$
& $\langle x^n\rangle_u^{\mathrm{tot}}$ & $\langle x^n\rangle_u^{3q}$ & $\langle x^n\rangle_u^{3qg}$
& $\langle x^n\rangle_d^{\mathrm{tot}}$ & $\langle x^n\rangle_d^{3q}$ & $\langle x^n\rangle_d^{3qg}$
& $\langle x^n\rangle_g$ \\
\midrule
\multicolumn{8}{c}{$n=0$} \\
\midrule
0     & 2        & 0.837387 & 1.16261 & 1        & 0.418693 & 0.581307 & 0.581307 \\
0.662 & 2        & 0.775632 & 1.22437 & 1        & 0.387816 & 0.612184 & 0.612184 \\
1.0  & 2        & 0.748881 & 1.25112 & 1        & 0.37444 & 0.62556 & 0.62556 \\
1.583 & 2        & 0.710601 & 1.28940 & 1        & 0.355301 & 0.644699 & 0.644699 \\
\midrule
\multicolumn{8}{c}{$n=1$} \\
\midrule
0     & 0.595429 & 0.288621 & 0.306808 & 0.255481 & 0.130073 & 0.125409 & 0.149090 \\
0.662 & 0.585543 & 0.266902 & 0.318641 & 0.250662 & 0.120914 & 0.129748 & 0.163795 \\
1.0 & 0.580413 & 0.257322 & 0.32309 & 0.248626 & 0.117118 & 0.131508 & 0.170962 \\
1.583 & 0.571845 & 0.243345 & 0.328501 & 0.245929 & 0.111956 & 0.133973 & 0.182225 \\
\bottomrule
\end{tabular}
\end{minipage}
\end{table*}

Further insight into the EOS is obtained by examining the pressure as a function 
of the nuclear density. Figure~\ref{densi} compares the pressure predicted by 
our QMC-BLFQ model with several theoretical models as well as empirical constraints 
extracted from heavy-ion collisions. The band from Ref.~\cite{Lynch2009} represents 
constraints derived from the analyses of nuclear reactions in a transport model. The 
QMC-BLFQ results fall within this band over a broad range of densities, indicating 
that the model predicts a realistic stiffness of the nuclear matter EOS.

The pressure constraint extracted from the flow  analysis of heavy-ion collisions reported in Ref.~\cite{Danielewicz:2002pu} is shown by the band in Fig.~\ref{densi} (bottom left), where
the QMC-BLFQ prediction  remains largely compatible with this constraint up to 
moderate densities. In addition, QMC-BLFQ is compatible with QHD models, MQMC, MHD and the band representing the data extracted from \cite{Huth2021} and references therein (bottom right). Overall, the QMC-BLFQ framework provides a consistent description of the EOS's for SNM, while explicitly incorporating quark and gluon degrees of freedom.

\section{In-medium nucleon parton distributions}\label{sec:PDF_SNM}

One of the main motivations for the present framework is the study of the
modification of parton distributions in the nuclear medium.
Since the nucleon structure in the QMC-BLFQ approach is determined at the
quark level, the density dependence of the scalar and vector mean fields
induces a modification of the internal LFWFs of the
nucleon. Then, the in-medium nucleon wave function  is:
\begin{equation}
\Psi^{\Lambda*}_{\mathcal{N}, \lbrace \lambda_{i} \rbrace}(\{x_i, \mathbf{p}_{i\perp}\}) = \sum_{\{n_i, m_i\}} \psi^{\Lambda*}_\mathcal{N}(\{\alpha_i\}) \prod_i^\mathcal{N} \phi_{n_i, m_i}(\mathbf{p}_{i\perp}; b)\,,
\end{equation}
where the star indicates the  dependence on the nuclear density.

As a consequence, the PDFs acquire
an explicit dependence on the nuclear density. At leading twist, the in-medium quark and gluon PDFs are written as
\begin{equation}
f^{q,g}_{\rho/\rho_0}(x)
=
\sum_{\mathcal N,\{\lambda_i\}}
\int [d\Gamma_\mathcal N]\,
\left|
\Psi^{\Lambda *}_{\mathcal{N},\{\lambda_i\}}
(\{x_i,\mathbf{p}_{i\perp}\})
\right|^2
\delta(x-x_j) ,
\end{equation}
where the integration measure is defined as
\begin{equation}
[d\Gamma_\mathcal N]
=
\prod_{i=1}^{\mathcal N}
dx_i\, d^2\mathbf p_{i\perp}\,
\delta\!\left(1-\sum_{i=1}^{\mathcal N} x_i\right)
\delta^{(2)}\!\left(\sum_{i=1}^{\mathcal N} \mathbf p_{i\perp}\right),
\end{equation}
with $j$ denoting the active parton ($q$: quark and $g$: gluon). In the present nucleon model, the quark PDF $f^\ast_{q}(x)$ receives contributions from both $|qqq\rangle$ and $|qqqg\rangle$ components of the LFWFs, while the gluon distribution is obtained only from the $|qqqg\rangle$ 
component.

The medium modification of the PDFs can be directly related to the change in the
effective quark masses and consequently the internal structure of the nucleon evaluated with BLFQ, as an eigenstate of the in-medium LF Hamiltonian~\eqref{eq:P-*}.
Therefore, the medium effect on the nucleon PDFs leads to a redistribution of the
LF momentum fractions carried by the quarks and gluons. In Table~\ref{tab:pdf_moments}, the effects of nuclear density on the probabilities (zeroth moment, $\langle x^0\rangle$) and the first moment ($\langle x^1\rangle$) of the parton distributions are shown at the model scale $\mu_0^2 = 0.291~\rm{GeV}^2$~\cite{Xu:2022yxb}.
With increasing density, the gluon probability, $\langle x^0\rangle$, for the $|qqqg\rangle$ component raises, { and at the saturation density there is a 10\%  enhancement.} This enhancement is primarily driven by medium-induced modifications of the quark--gluon dynamics and LFWFs.

Table~\ref{tab:pdf_moments} presents the momentum fractions carried by each constituent in the $|qqq\rangle$ and $|qqqg\rangle$ components of the nucleon state.
The quark momentum fraction in the valence component roughly decreases with the valence probability, while the quark momentum fraction in the $|qqqg\rangle$ component increases as its probability grows. The gluon momentum fraction rises up to 20\% for $\rho/\rho_0 = 1.583$, whereas $\langle x^0 \rangle_g$ increases by only 10\%, showing a different trend compared to the quark momentum fractions in both Fock components.

\begin{figure}[thb]
    \centering
    \includegraphics[width=0.99\linewidth]{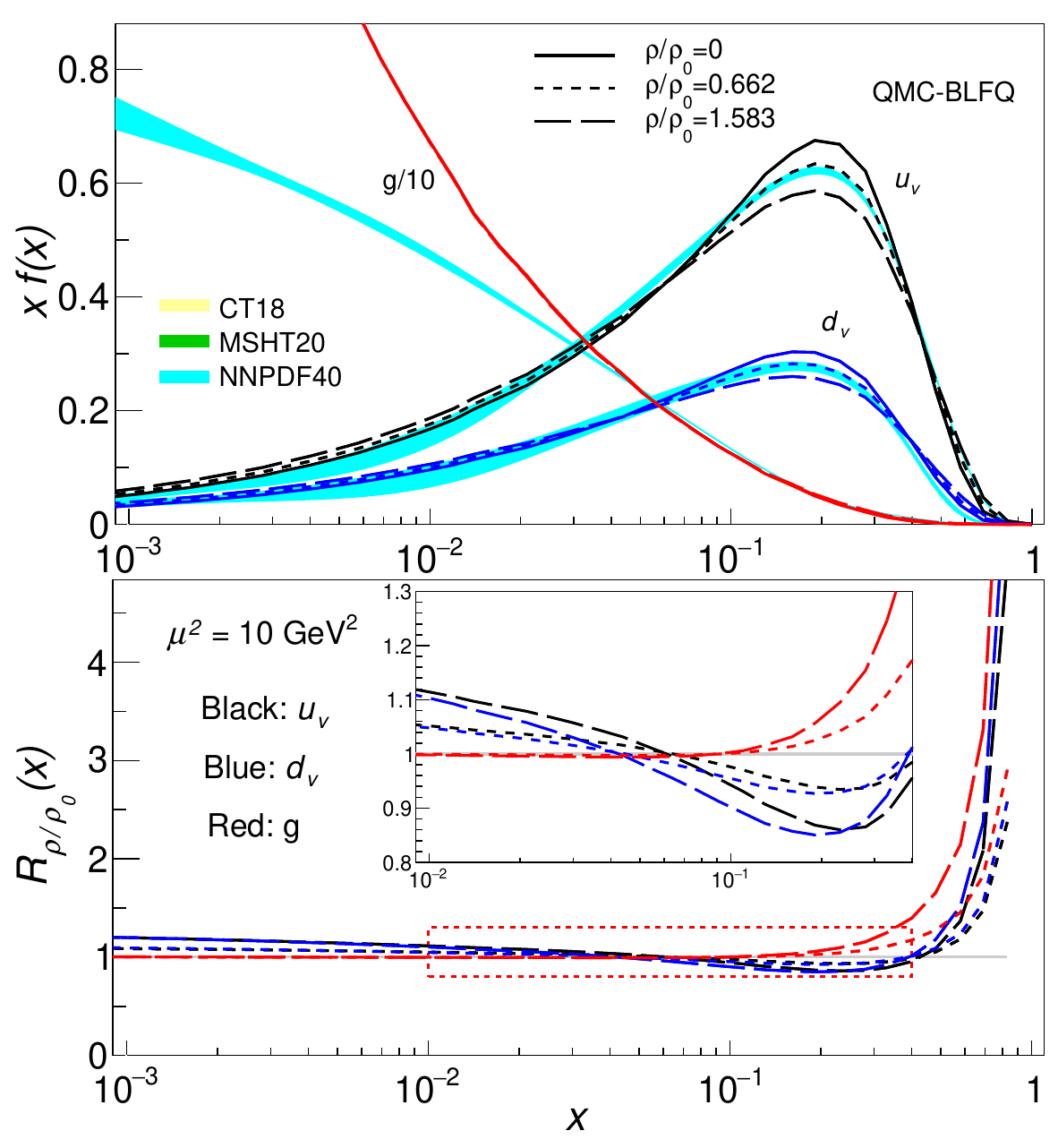}
    
\caption{PDFs of the proton obtained within the QMC-BLFQ framework at the scale $\mu^2 = 10~\mathrm{GeV}^2$ for different nuclear densities $\rho/\rho_0$. The upper panel shows $x~f^i(x)$ for $u_v$ (black), $d_v$ (blue), and gluon $g$ (red), compared with global fits from CT18, MSHT20, and NNPDF4.0 from LHAPDF6~\cite{Buckley:2014ana}. The lower panel displays the ratios $R_{\rho/\rho_0}^i(x)$, defined in Eq.~(\ref{ratioEq}), for $i=u_v, d_v, g$, illustrating the density dependence of the parton distributions. The inset shows a magnified view of the intermediate-$x$ region highlighted by the red dashed rectangle in the main panel.}
    \label{fig:xpdf}
\end{figure}
\begin{figure}[thb]
    \centering
    \includegraphics[width=0.99\linewidth]{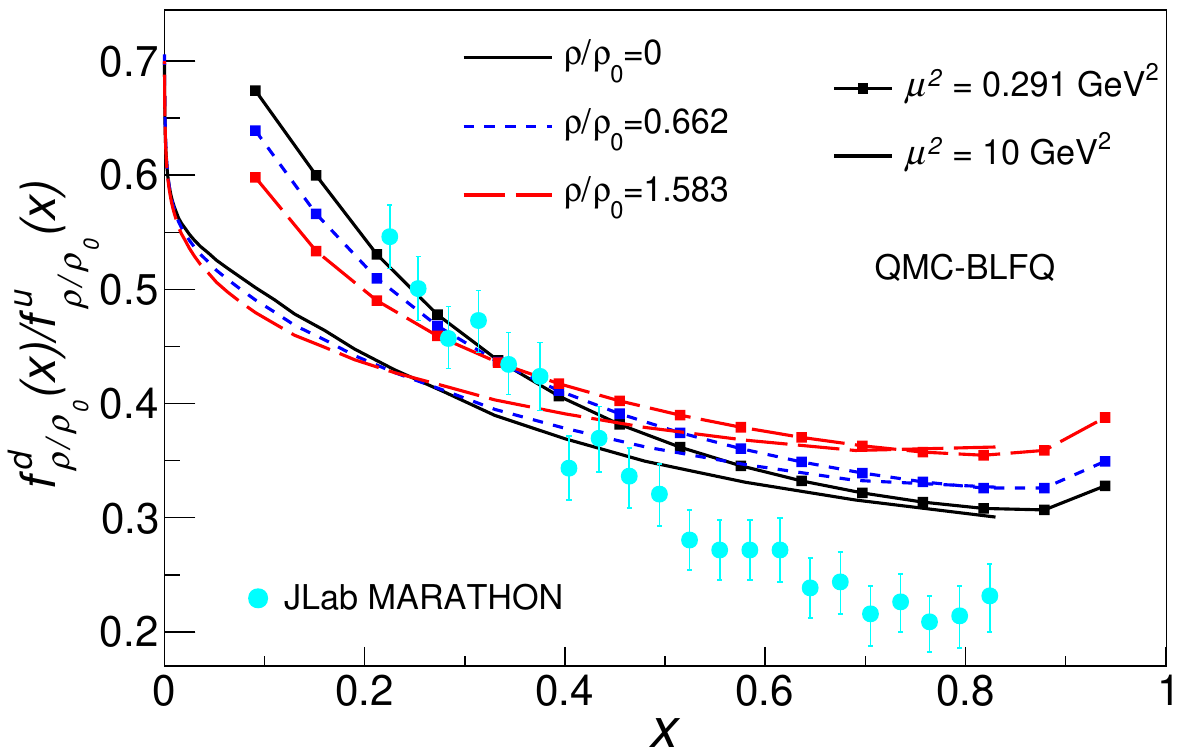}
    \caption{Proton $f^d_{\rho/\rho_0}(x)/f^u_{\rho/\rho_0}(x)$ for $\mu_0^2=0.291\,$GeV$^2$ (lines with marker) and $\mu^2=10\,$GeV$^2$ (lines without marker) with different nuclear densities: $\rho/\rho_0=0$ (solid-black line), $\rho/\rho_0=0.662$ (dashed-blue line), $\rho/\rho_0=1.583$  (long-dashed-red line). The cyan data are the  extracted data from the Jefferson Lab MARATHON experiment \cite{JeffersonLabHallATritium:2021usd}.}
    \label{fig:doveru}
\end{figure}

The parton distributions $x\,f(x)$ evolved to $\mu^2=10\,$GeV$^2$ are presented in Fig.~\ref{fig:xpdf} (top) for the quarks and gluon, in vacuum and for different densities. The $x\,f_{(u,d)}(x)$ is depleted at the peak by about at most  10\%  for $\rho/\rho_0=1.583$, for each flavor, as indicated already  by the decrease of the momentum fraction carried by the quarks. For the gluon $x\,f_{g}(x)$ the  difference between the in-medium and vacuum is difficult to observe in the top panel of the figure, but it mainly happens for large values of $x$, as it is  clear in the bottom panel.

Figure~\ref{fig:xpdf} (top) also shows the results for $xf_{u,d,g}(x)$ from different global analyses, namely CT18, MSHT20, and NNPDF4.0 from LHAPDF6~\cite{Buckley:2014ana}, at $\mu^2 = 10~\rm{GeV}^2$. The vacuum proton valence distributions $xu_v$ and $xd_v$ are close to the bands provided by these analyses, while the gluon distribution overestimates the global fits for $x \lesssim 0.04$. This suggests that the interaction coupling the two Fock components may be overestimated at small $x$, and/or reflects the missing gluon fusion dynamics in the standard Dokshitzer-Gribov-Lipatov-Altarelli-Parisi (DGLAP) evolution.

The ratio of the in-medium $f_{u,d}(x)$ to the vacuum distributions is shown in Fig.~\ref{fig:xpdf} (bottom). A slight enhancement above 1 is observed at small $x$, while for $x\sim0.3$ the ratio drops to about 0.85, and for $x \gtrsim 0.4$ it rises consistently above unity. This suppression of the quark distribution at intermediate $x$ is a well-known feature referred to as the EMC effect~\cite{Geesaman:1995yd,Norton:2003cb}. The in-medium gluon distribution, $xf_g(x)$, remains close to 1 up to $x \sim 0.1$, and for $x \gtrsim 0.1$ it shows a pronounced increase, which becomes more significant at higher nuclear densities.

The enhancement at large $x$ of the ratios
\begin{equation}
\label{ratioEq}
R_{\rho/\rho_0}^i(x)=\frac{f_{\rho/\rho_0}^i(x)}{f_0^i(x)} \,,
\qquad i=q,\,g ,
\end{equation}
is a direct consequence of in-medium effects at the mean-field level. Our model does not include nucleon overlap effects arising from the nucleon-nucleon short-range correlation (SRC) and the associated modifications of the nucleon structure, which remain an important topic of current interest (see, e.g.,~\cite{nCTEQ:2023cpo,Fomin:2026swt}).

Figure~\ref{fig:doveru} presents the ratio of down to up quark PDFs in the proton, $f^d_{\rho/\rho_0}(x)/f^u_{\rho/\rho_0}(x)$, as a function of $x$. The results are obtained within the QMC-BLFQ framework at three nuclear densities: $\rho/\rho_0 = 0$ (vacuum), $\rho/\rho_0 = 0.662$ (moderate), and $\rho/\rho_0 = 1.583$ (high). Calculations are shown both at the initial model scale, $\mu_0^2 = 0.291~\mathrm{GeV}^2$, and after QCD evolution to $\mu^2 = 10~\mathrm{GeV}^2$. Comparison with the JLab MARATHON experimental data~\cite{JeffersonLabHallATritium:2021usd} demonstrates that the QMC-BLFQ predictions capture the overall trend of the ratio. After evolution to $10~\rm{GeV}^2$, nuclear density enhances the ratio in the large-$x$ region ($x \gtrsim 0.5$) by 10--20\% for the densities considered. 

Evolution to higher scales also shifts the distribution toward lower $x$ and smooths the curves, consistent with standard DGLAP evolution. These results highlight the sensitivity of the proton's quark structure to nuclear medium effects and confirm the effectiveness of the QMC-BLFQ approach in describing medium-modified PDFs.

\section{Summary}\label{sec:summary}

The in-medium quarks couple to scalar and vector fields treated at the mean-field level: the scalar field modifies the effective quark mass, while the vector field induces a shift in the light-front energy. These effects are incorporated into the nucleon in-medium light-front Hamiltonian. Using BLFQ, the in-medium nucleon effective mass and wave function are computed as functions of the scalar mean field, introducing a density dependence after solving the self-consistent equation for the scalar field. The model has only two free parameters, the scalar coupling $g^q_{\sigma}$ and the vector coupling $g^q_{\omega}/m_\omega$, fitted to the energy per nucleon of symmetric nuclear matter at saturation. The resulting bulk properties at zero temperature, including energy per nucleon, pressure, and incompressibility, exhibit realistic saturation behavior in good agreement with experimental and empirical data~\cite{Huth2021,LeFevre2015,Russotto2016,Lynch2009,Danielewicz:2002pu}.

In the in-medium nucleon, the LFWFs exhibit a small increase in the gluon probability at saturation, accompanied by a corresponding decrease in the valence probability and in the momentum fraction carried by quarks. The in-medium unpolarized quark and gluon distributions show a noticeable enhancement for $x \gtrsim 0.4$ at the mean-field level, as illustrated at the evolved scale of $10~\mathrm{GeV}^2$. Furthermore, the proton $d/u$ ratio of the longitudinal momentum distributions for $x \gtrsim 0.4$ is increased compared with the free nucleon.

The present QMC-BLFQ model includes only mean-field effects in the nucleon structure and neglects nucleon–nucleon overlap, which gives rise to SRCs that distort the nucleon even in the deuteron. 
A complete description of nuclear PDFs would ideally incorporate both mean-field effects and SRCs. We also plan to extend the QMC-BLFQ model to finite temperatures and densities to study the phase diagram of symmetric and asymmetric nuclear matter. For asymmetric matter, this requires coupling quarks to the isovector vector mean field, as in neutron-star matter, and ensuring consistency with observational constraints~\cite{MALIK2025100086}, enabling exploration of compact star properties within the QMC-BLFQ framework.

\section*{Acknowledgments}
This work is supported by the National Natural Science Foundation of China under Grant No. 12305095 and No.12375143, by the Gansu International Collaboration and Talents Recruitment Base of Particle Physics (2023-2027), by the Senior Scientist Program funded by Gansu Province, Grant No. 25RCKA008.
JL is supported by the Special Research Assistant Funding Project, Chinese Academy of Sciences, and by Gansu Provincial Young Talents Program. SK is supported by China Postdoctoral Science Foundation (CPSF), Grant No. E339951SR0. XZ is supported by Key Research Program of Frontier Sciences, Chinese Academy of Sciences, Grant No. ZDBS-LY-7020, by international partnership program of the Chinese Academy of Sciences, Grant No. 016GJHZ2022103FN, by the Strategic Priority Research Program of the Chinese Academy of Sciences, Grant No. XDB34000000.
JPBCM and TF thank the hospitality of the Institute of Modern Physics (IMP),  to the  President’s International Fellowship Initiative/ Chinese Academy of Sciences (PIFI/CAS) (Grant No. 2026PVA0083, 2026PVA0216) for the financial support during their visit to IMP,  to Instituto Nacional de Ci\^encia e Tecnologia - Nuclear Physics and Applications (INCT-FNA, MCTI) (Grant No. 408419/2024-5), and to Funda\c c\~ao de Amparo \`a Pesquisa do Estado de S\~ao Paulo (FAPESP) (Grant No. 2024/17816-8). JPBCM thanks the support from FAPESP (Grant
No. 2023/09539-1) and to the Conselho Nacional de Desenvolvimento Científico e Tecnológico (CNPq) (Grant No. 351403/2025-6). TF thanks the support from CNPq (Grant No. 306834/2022-7) and FAPESP (Grant No. 2023/13749-1 and 2025/05312-8).

\biboptions{sort&compress}
\bibliographystyle{elsarticle-num}
\bibliography{references}

\end{document}